# An Application of a Modified Beta Factor Method for the Analysis of Software Common Cause Failures

**Tate Shorthill[a], Han Bao[b], Edward Chen[c], and Heng Ban[d]**
[a] University of Pittsburgh, Pittsburgh, PA, USA, ths60@pitt.edu
[b] Idaho National Laboratory, Idaho Falls, ID, USA, han.bao@inl.gov
[c] North Carolina State University, Raleigh, NC, USA, echen2@ncsu.edu
[d] University of Pittsburgh, Pittsburgh, PA, USA, heng.ban@pitt.edu

**Abstract**: This paper presents an approach for modeling software common cause failures (CCFs) within digital instrumentation and control (I&C) systems. CCFs consist of a concurrent failure between two or more components due to a shared failure cause and coupling mechanism. This work emphasizes the importance of identifying software-centric attributes related to the coupling mechanisms necessary for simultaneous failures of redundant software components. The groups of components that share coupling mechanisms are called common cause component groups (CCCGs). Most CCF models rely on operational data as the basis for establishing CCCG parameters and predicting CCFs. This work is motivated by two primary concerns: (1) a lack of operational and CCF data for estimating software CCF model parameters; and (2) the need to model single components as part of multiple CCCGs simultaneously. A hybrid approach was developed to account for these concerns by leveraging existing techniques: a modified beta factor model allows single components to be placed within multiple CCCGs, while a second technique provides software-specific model parameters for each CCCG. This hybrid approach provides a means to overcome the limitations of conventional methods while offering support for design decisions under the limited data scenario.

## 1. INTRODUCTION

Digital instrumentation and control (I&C) systems offer many benefits over their traditional analog counterparts; however, technical challenges and costs associated with ensuring their safe and reliable implementation have slowed the adoption of digital upgrades within the nuclear industry [1]. In 1997, the United States (U.S.) Nuclear Regulatory Commission funded research to identify the challenges of implementing digital I&C systems within the nuclear industry [2]. The identification, quantification, prevention, and mitigation of potential common cause failures (CCFs) within digital I&C systems remains a relevant technical challenge today [3]. This work presents a approach for CCF analysis as part of the Idaho National Laboratory (INL) framework for the risk assessment of digital I&C systems developed under the Risk-Informed Systems Analysis (RISA) Pathway of the U.S. Department of Energy (DOE) Light Water Reactor Sustainability (LWRS) program [4, 5, 6].

A CCF is the occurrence of two or more failure events due to the simultaneous occurrence of a shared failure cause and a coupling factor (or mechanism) [7]. The failure cause is the condition to which failure is attributed, whereas the coupling mechanism creates the condition for the failure cause to affect multiple components, thereby producing a CCF [7]. Some examples of coupling mechanisms given in NUREG/CR-5485 include design, hardware, function, installation, maintenance, and environmental conditions [7]. Any group of components that share similarities via coupling mechanisms may have a vulnerability to CCF; a group of such components are considered a common cause component group (CCCG) [7]. The identification of coupling factors and, by extension, CCCGs is an essential part of CCF analysis. Often, CCF models attempt to simplify an analysis by assuming symmetry for the components of a CCCG. For example, a CCCG may be assigned by assuming components are identical where any differences in the coupling factors are ignored. There are many methods for modeling CCFs, including direct assessment methods, ratio models (e.g., beta factor and alpha factor models), Bayesian inference methods, and shock models [8]. Nearly all of them rely on symmetry; the most notable exceptions are the direct assessment methods and those based on Bayesian inference. However, it may be important to explicitly consider the influences of multiple coupling factors that might otherwise be ignored by the symmetry assumption. A software failure is the direct result of operational conditions (i.e., a trigger scenario) activating some hidden software defect(s) causing the inability of the software to perform its require or intended functions (based on concepts from [9] and [10]). A software CCF will occur when a coupling mechanism creates a scenario for operational conditions to activate a common software defect. Given a group of redundant software components, variations in their operating conditions may lead to some, but not all, components failing together. Variation of maintenance activities, input variable sources, component locations, and installation teams influence the operational environment; ultimately, subtle differences in coupling mechanisms may influence which components fail together. Capturing asymmetry between components may be necessary for software CCF modeling, but it can be challenging with conventional methods.

**Figure 1. Example system showing the relationship of independent and dependent failures in the context of a fault tree.**

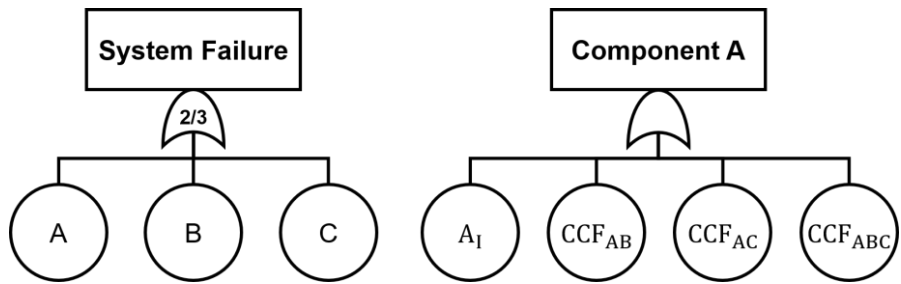

Consider a scenario shown in Figure 1 where the components are arranged in the 2/3 criteria for failure. The probability of failure for the system, as given in NUREG/CR-5485, is shown below:

$$P(F) = P(A_I)P(B_I) + P(B_I)P(C_I) + P(A_I)P(C_I)$$
$$+ P(CCF_{AB}) + P(CCF_{BC}) + P(CCF_{AC}) + P(CCF_{ABC}) \quad (1)$$

The common practice in reliability modeling is to assume the failure probabilities (or rates) of similar components are the same [7]. This symmetry assumption results in the following:

$$Q_1^3 = P(A_I) = P(B_I) = P(C_I) = Q_1 \quad (2)$$

$$Q_2^3 = P(CCF_{AB}) = P(CCF_{BC}) = P(CCF_{AC}) = Q_2 \quad (3)$$

$$Q_3^3 = P(CCF_{ABC}) = Q_3 \quad (4)$$

where $Q_k^m$ represents the failure rate or probability of an event involving k components in a CCCG of size m. Now, consider the case when the components of Figure 1 share some, but not all, coupling factors. In this new scenario, components A, B, and C are coupled by procedures, while A and B are coupled by location. The options are to either ignore the differences or to account them directly. Ignoring the differences leads to a single CCCG and reliance on Equations 1-4. When the differences are considered, the traditional approach forms two CCCGs: (1) CCCG1 with components A, B, and C; and (2) CCCG2 with components A and B. This ultimately requires a single component to be part of multiple CCCGs. The symmetry assumption applied to CCCG1 relies on the same equations as above. The symmetry assumption applied to CCCG2 gives:

$$Q_1^2 = P(A_I) = P(B_I) = Q_1 \quad (5)$$

$$Q_2^2 = P(CCF_{AB}) = Q_2 \quad (6)$$

Placing A and B within multiple CCCGs creates additional challenges because conventional models (i.e., the alpha factor model [7]) may provide two different probabilities for the same CCF event [11].

For example, some CCF models may determine $P(CCF_{AB})$ from CCCG1 to be different than $P(CCF_{AB})$ from CCCG2. This is because conventional models incorporate the CCCG size as part of their evaluation process and consider combinations of failures between the components of the CCCG. If modeling is performed using a program such as the Systems Analysis Programs for Hands-on Integrated Reliability Evaluations (SAPHIRE) [12], having a single component within multiple CCCGs may lead to double counting of failure events. Ma et al. address this issue further and suggest using the largest CCCG that is reasonable [11]. However, this solution requires the analyst to ignore the potential asymmetry of the coupling factors. They suggest a second option may be to select which value of the duplicate failure events is appropriate. Additional examples exist that allow components to be part of multiple CCCGs, such as when each CCCG represents a unique failure mode [11].

In order to directly consider subtle differences in coupling mechanisms, there are two approaches for forming the CCCGs. The first, as mentioned earlier, forms unique CCCGs for each shared set of coupling factors and may require some components to be part of multiple CCCGs. The second approach forms CCCGs that allow for some variation in the coupling mechanisms (e.g., from Figure 1, a single CCCG that contains A, B, and C, but allows for differences in $P(CCF_{AB}), P(CCF_{BC})$, and $P(CCF_{AC})$ directly, rather than assume they all equal $P(Q_2)$). The second approach requires an asymmetric model to directly account for these differences within the CCCG. Models for asymmetry and models that allow components to be part of multiple CCCGs have been addressed by several publications. Rasmussen and Kelly proposed a method to deal with asymmetric failure probabilities within the context of the basic parameter model [13]. In 2012, Kančev and Čepin proposed a modification of the beta factor model that allows components to be assigned to multiple CCCGs based on their coupling factors [14]. O'Connor and Mosleh proposed a partial alpha factor model and a Bayesian approach (the general dependency model); an extension to the alpha factor model, the partial alpha factor works to explicitly model coupling factors between components [15]. The general dependency model relies on a Bayesian network to account for three parameters—a cause condition probability, component fragility, and coupling factor strength [16]. In 2020, Higo et al. developed a method to account for the combined influence of asymmetric and symmetric CCF probabilities by assessing the degree of shared coupling factors [17]. This work was later refined by combining with a gamma factor model to express inter-unit CCF probability [18]. The challenge with these methods is their dependence on proprietary data for model parameters. Far less data is available for software-based CCFs, than for analog CCFs which challenges the application of these recent innovations. In addition, those methods that account for qualitative differences in coupling mechanisms (e.g., [8] and [16]) rely on data that may not exist for newly designed software systems. The goal of our work is to quantify software CCFs given minimal data while also considering the influence of software attributes on coupling mechanisms. Given most asymmetric models require data that is unavailable for software, we will forgo the formation of asymmetric CCCGs and instead rely on approach that considers qualitative information for CCF modeling while also allowing components to be part of multiple CCCGs.

This work proposes an approach for modeling software CCF given: (1) a lack of operational and CCF data for defining software CCF model parameters; and (2) the need to model single components as part of multiple CCCGs simultaneously. The model best suited for a limited data scenario may be the one requiring the fewest parameters. In this case, the modified beta factor model by Kančev and Čepin (referred to as the modified BFM in our work) is demonstrated for software CCF analysis. Section 2 details our methodology for modeling software CCF including innovations for defining software-specific model parameters. Section 3 provides a case study. Finally, Section 4 discusses our results and conclusion.

## 2. METHODOLOGY

This section is focused on answering two needs for modeling software CCFs. The first part of this section discusses an approach for modeling components as part of multiple CCCGs simultaneously as provided by the modified BFM. The second half details the innovative application of the modified BFM for software CCF analysis. Specifically, this section details our innovations for addressing the lack of operational and CCF data typically used to define model parameters.

The modified BFM, as its name suggests, is based on the beta factor model [14]. The beta factor model is one of the oldest CCF models and assumes that a total failure probability ($Q_t$) of a component is a contribution of independent ($Q_I$) and dependent ($Q_D$) failures; the dependent failure probability is given as a fraction (i.e., $\beta$) of the total failure probability ($Q_t$) of the component as observed in Equation (7)). Likewise, the independent failure is also a function of $\beta$. The beta factor model implements the symmetry assumption such that all the components within a CCCG fail together according to the dependent (i.e., CCF) probability defined by beta. The model does not account for combinations of failures within a CCCG [7]. The beta factor model applied to a CCCG of A, B, and C will only find $CCF_{ABC}$. Therefore, the only way to consider a CCF of two components is to assign them their own CCCG. This is the basis of the modified BFM. Our work assumes that the potential for combinations of failures with the CCCG is largely dependent on the existence of subtle differences in the coupling mechanisms. Hence, to account for any distinct CCFs, we rely on coupling factor-based CCCGs.

$$Q_t = Q_I + Q_D \tag{7}$$

$$Q_D = \beta Q_t \tag{8}$$

$$Q_I = (1-\beta)Q_t \tag{9}$$

The modified BFM is designed to allow components to be members of multiple CCCGs [14]. Like the beta factor model, the modified BFM assumes the total failure probability/rate of a component is the summation of independent and dependent failures. Equation (10) shows the basis of the modified BFM, which is that the total dependent failure consists of the contribution of each CCCG failure. Each CCCG is assigned a group beta ($\beta_w$) that represents the contribution of that CCCG to the total failure probability. Equation (14) shows the independent failure probability in terms of each CCCG beta and total failure probability.

$$Q_D = P(CCCG_1) + P(CCCG_2) + \cdots P(CCCG_w) \tag{10}$$

$$P(CCCG_w) = (\beta_w)Q_t \tag{11}$$

$$\beta_t = \sum_{1}^{w}(\beta_w) \tag{12}$$

$$Q_D = Q_t \sum_{1}^{w}(\beta_w) \tag{13}$$

$$Q_I = (1-\beta_t)Q_t = \left[1 - \sum_{1}^{w}(\beta_w)\right]Q_t \tag{14}$$

Some advantages of this method include its ease of application, its consideration of CCCG-specific coupling factors, and its ability to account for multiple CCCGs directly. Double counting is avoided because the model assumes that CCFs represent the failure of each component within the CCCG and no other sub-combinations. For example, given two CCCGs (e.g., components A, B, and C for CCCG1 and A and B for CCCG2), there will be no chance of counting $P(CCF_{AB})$ twice because $P(CCF_{AB})$ is only evaluated for CCCG2. The modified BFM, like most methods, requires reference data to determine each CCCG failure probability/rate. Like other ratio models, the quantification of its parameters can be challenging for a limited-data scenario. The modified BFM is limited to identical components with identical total failure probabilities. If the $Q_t$ for the components within a CCCG are not identical, depending on the $Q_t$ selected for Equation (11), there will be differing values for the same CCFs. Sources [13] and [19] provide support for this scenario. An additional limitation can occur if the total beta, shown by Equation (12), exceeds unity. If this happens, then the summation of dependent failures will exceed the total failure probability. To account for this issue, Kančev and Čepin indicate a possible solution is to normalize the CCCG beta factors such that they sum to unity while maintaining their relative magnitudes. The second and third options include normalizing by the largest CCCG beta or using weight factors for each CCCG, respectively [14]. It is best to select the option which matches model assumptions (e.g., the first option will work better for software CCF low diversity systems, because it is expected that dependent software failure will exceed the independent software failure probability). Despite its known limitations, this work will employ the modified BFM for the quantification of CCFs because it works directly for the multiple CCCG scenario.

The next challenge is defining the model parameters. The emphasis of the current work is the limited-data scenario that naturally requires some form of expert elicitation. For elicitation, it is desirable to consider qualitative defenses against CCFs [19, 20]. There are at least two methods presented in literature that express the elicitation of the beta parameter without the use or dependence on operational data. These two methods, both of which are called "partial beta methods," develop beta from a combination of partial attributes; one employs an additive scheme to find beta [19], while the other a multiplicative scheme [20].

The first method, called partial beta factor-1 (PBF-1) in our work, was developed on the claim that dependent failures could not be determined without an engineering assessment of that system's defenses for such failures [20]. An assessment is made according to 19 defenses (e.g., functional diversity, maintenance, etc.), where each defense receives a partial beta value (i.e., $\beta_i$ between zero and one, where a zero score indicates a high defense against CCF). The product of the 19 scores is then used as the beta factor for the system. This multiplicative scheme may tend to predict small values for beta. For example, if 18 of the defenses are given $\beta_i = .99$, the CCF likelihood for the system should be high. However, the remaining defense ($\beta_{19}$) can dominate the system, resulting in an improper score for the system beta (e.g., if $\beta_{19} = .1$ and $\beta_{1-18} = .99$, then $\beta = .083$). Further complications could arise if additional defense categories are added. Ultimately, PBF-1 may underpredict dependent failures.

The second method, called partial beta factor-2 (PBF-2), does not actually use partial betas, rather the method uses a collection of sub-factors that contribute to an overall beta score [19]. Humphreys' method was later modified by Brand [21] and served as a foundation for a hardware CCF model used in the International Electrotechnical Commission (IEC) 61508 [22]. The PBF-2 was founded on the question, "What attributes of a system reduce CCFs?" [19]. These attributes, called sub-factors, are shown in Table 1. Each sub-factor was weighted by reliability engineers for their importance. The methodology requires the analyst to assign a score (e.g., A, B, C, etc.) for each sub-factor. An "E" indicates a component is well-defended against CCFs (i.e., A= poor, E= ideal). The sub-factor names alone are not sufficient for assessing each sub-factor; therefore, readers are advised to visit the original source material for scoring guidance. Beta, given by Equation (15), is a function of the assigned sub-factor scores and the denominator $d$. The model was arranged such that the upper and lower limits for beta correspond with dependent failure values reported in literature [19]. The limits are ensured by the sub-factors and $d$ given in Table 1. The beta value determined by this method was intended to be used with beta factor model; but in this work, it will be used with the modified BFM.

$$\beta = \frac{\sum(Sub-factors)}{d} \qquad (15)$$

PBF-2 provides a convenient and structured determination of beta associated with the hardware failure of digital I&C components, yet only minimal consideration is provided for software [19]. In fact, some methods (e.g., IEC 61508) prefer to provide qualitative approaches to avoid or control software failures [23]. In contrast, this work emphasizes the quantification of both hardware and software failures. As mentioned, CCFs are conditional on a shared root cause and coupling factor. Within the context highly redundant digital I&C systems, and low instances of software diversity, it is anticipated that CCFs should represent a significant portion of the software failure. Redundant components share application software failure by nature of their common (i.e., identical) software.

Software failure occurs by the activation of latent defects (e.g., deficiencies from coding errors, installation errors, maintenance errors, setpoint changes, requirements errors, etc.). Activation of latent defects is a result of certain operational conditions (i.e., trigger events) [10]. Trigger events act as software inputs, without which there would be no fault activation and, ultimately, no failure. A software CCF will result from a shared root cause (i.e., a shared trigger event and a defect) leading to the failure of two or more components by means of a coupling mechanism. Coupling mechanisms influence how a trigger event and/or a defect is shared by multiple components. As an example, consider that a software developer (i.e., a coupling mechanism) introduces a shared defect in redundant controllers allowing a trigger event to cause a CCF. In contrast, a maintenance procedure (i.e., a coupling mechanism) may shuts down half of a system thereby creating a condition for a trigger event to affect only the active components. Given a group of redundant software components, variations in their operating conditions may lead to some, but not all, components failing together. Variations in the operational environment of otherwise identical components may result from differences in maintenance staff, inputs variables, etc. In other words, subtle differences in coupling mechanisms may lead to unique combinations of CCFs. Thus, it is essential to consider software-based coupling mechanisms when assessing the potential for CCFs within a digital I&C system. To account for software features, PBF-2 was modified in two ways: (1) the model was adjusted to increase the upper and lower limits of beta (i.e., 0.001 – 0.999), allowing for greater applicability to low diversity software systems; and (2) the sub-factor weights were changed to emphasize software-centric features. It is understood that diversity affects CCFs [10]. Consequently, the sub-factors that influence diversity were weighted heavily. As an example, the adjusted model emphasizes the introduction of software faults and coupling mechanisms by placing greater weight on those defenses that pertain to human interaction and the diversity of software. Subtle variations in the coupling mechanisms create quasi-diverse components, ultimately influencing the potential for CCFs. Table 2 shows the adjustments made to PBF-2 to account for software. It, along with Table 1, are used to define the beta factors for software and hardware failures, respectively. Sub-factors are scored according to the guidance given by [21] with some additional considerations for software: (1) to score Redundancy (& Diversity), the diversity is assessed (e.g., A indicates no diversity, while E indicates complete software diversity for the CCCG); (2) the testing category considers software operational testing; and (3) the separation category was changed to Input Similarity. Physical separation alone does not influence software failure unless there is consideration for how that physical separation changes the operational conditions of the components. Whereas the Redundancy (& Diversity) sub-factor considers the degree of internal similarity, the Input Similarity sub-factor considers the degree to which redundant software share external and input similarity. Guidance for scoring the Input Similarity is shown in Table 3.

**Table 1. Beta Factor Estimation Table for Hardware.**

| Sub-factors | A | A+ | B | B+ | C | D | E |
|---|---|---|---|---|---|---|---|
| Redundancy (& Diversity) | 1800 | 882 | 433 | 212 | 104 | 25 | 6 |
| Separation | 2400 | | 577 | | 139 | 33 | 8 |
| Understanding | 1800 | | 433 | | 104 | 25 | 6 |
| Analysis | 1800 | | 433 | | 104 | 25 | 6 |
| MMI | 3000 | | 721 | | 173 | 42 | 10 |
| Safety Culture | 1500 | | 360 | | 87 | 21 | 5 |
| Control | 1800 | | 433 | | 104 | 25 | 6 |
| Tests | 1200 | | 288 | | 69 | 17 | 4 |
| Denominator for Equation (15), $d = 51000$. |||||||||
| Note: The current work relies on an automatic calculation that provides slightly different table values than those given in the source material. The original derivation indicates that scoring an "A" for each sub-factor will result in 0.3 for the beta factor [19]. The current table provides 0.300 while the original provides 0.302. The difference is negligible, so this work employs the automated calculation for convenience. |||||||||

**Table 2: Beta Factor Estimation Table for Software**

| Sub-factors | A | A+ | B | B+ | C | D | E |
|---|---|---|---|---|---|---|---|
| Redundancy (& Diversity) | 23976 | 10112 | 4265 | 1799 | 759 | 135 | 24 |
| Input Similarity | 23976 | 10112 | 4265 | | 759 | 135 | 24 |
| Understanding | 7992 | | 1422 | | 253 | 45 | 8 |
| Analysis | 7992 | | 1422 | | 253 | 45 | 8 |
| MMI | 11988 | | 2132 | | 379 | 67 | 12 |
| Safety Culture | 6993 | | 1244 | | 221 | 39 | 7 |
| Control | 4995 | | 888 | | 158 | 28 | 5 |
| Tests | 11988 | | 2132 | | 379 | 67 | 12 |
| Denominator for Equation (15), $d = 100000$. |||||||||

Table 3. Sub-factor Guide for Input Similarity

| Score | R=0 | 0 < R < .5 | .5 ≤ R < 1 | R ≥ 1 | Zero Diversity | Partial Diversity | Complete Diversity |
|---|---|---|---|---|---|---|---|
| A | X | | | | X | X | X |
| A+ | | X | | | X | X | X |
| B | | | X | | X | | |
| C | | | X | | | X | X |
| D | | | | X | X | X | |
| E | | | | X | | | X |

The input ratio ($R$) is defined: $R = (s-1)/m$ for $s = 1$ and $R = s/m$ for $s > 1$ where, $m$ = the number of components within the CCCG, and $s$ = number of input sources.

This work presents an approach for performing CCF analysis on digital I&C systems given limited data by integrating the modified BFM and PBF-2. The approach relies on the modified BFM to allow components to be part of multiple CCCGs and PBF-2 defines beta factors for each CCCG. The hybrid approach provides a means to overcome limitations of conventional methods. A formalized process that relies on the modified BFM and PBF-2 is shown in Figure 2, which has been demonstrated in [24, 25]. The subsequent section will demonstrate this process as with a case study.

**Figure 2. Flowchart for Software CCF Modeling and Estimation.**

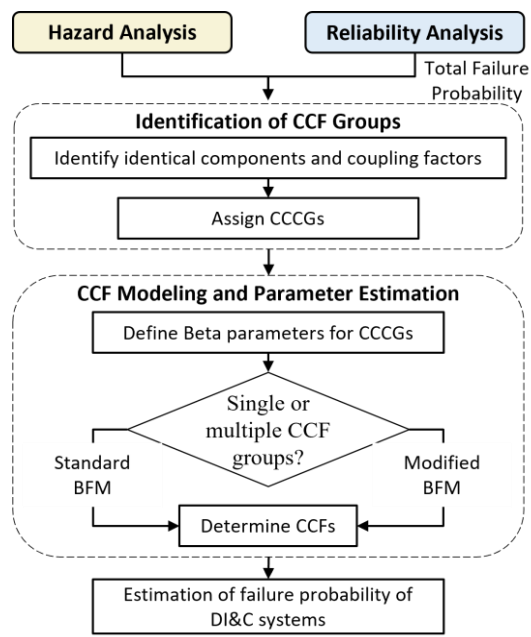

## 3. CASE STUDY

This case study describes the quantification of the CCFs found in the automatic trip function of a four-division digital reactor trip system (RTS). Division-based sensor signals are sent to the bistable processors (BPs), which determine whether a trip is needed. When required, trip signals from the BPs are sent to each of the divisions' local coincidence logic processors (LPs). The LPs vote on the incoming trip signals and send the output via digital output modules (DOMs) to selective relays, which again vote on the trip signals. The outputs of the selective relays pass through undervoltage trip devices (e.g., RTB-D1-UV) and activate the undervoltage reactor trip breakers (e.g., RTB-A1). The correct combination of breakers results in a reactor trip. Diverse trip mechanisms (e.g., shunt trip devices like RTB-DA-ST) via the diverse protection system (DPS) and manual trip mechanisms via the main control room (MCR) or the remote shutdown room (RSR) are not part of the case study. Table 4 provides the list of components for which failure rates need to be quantified. In this work, the only components shown in Figure 3 to contain application software are the BPs and LPs, both of which are programmable logic controllers. Evaluation of the software CCF values follows the approach described in the previous section.

**Figure 3. Four-Division Digital Reactor Trip System (adapted from [26]).**

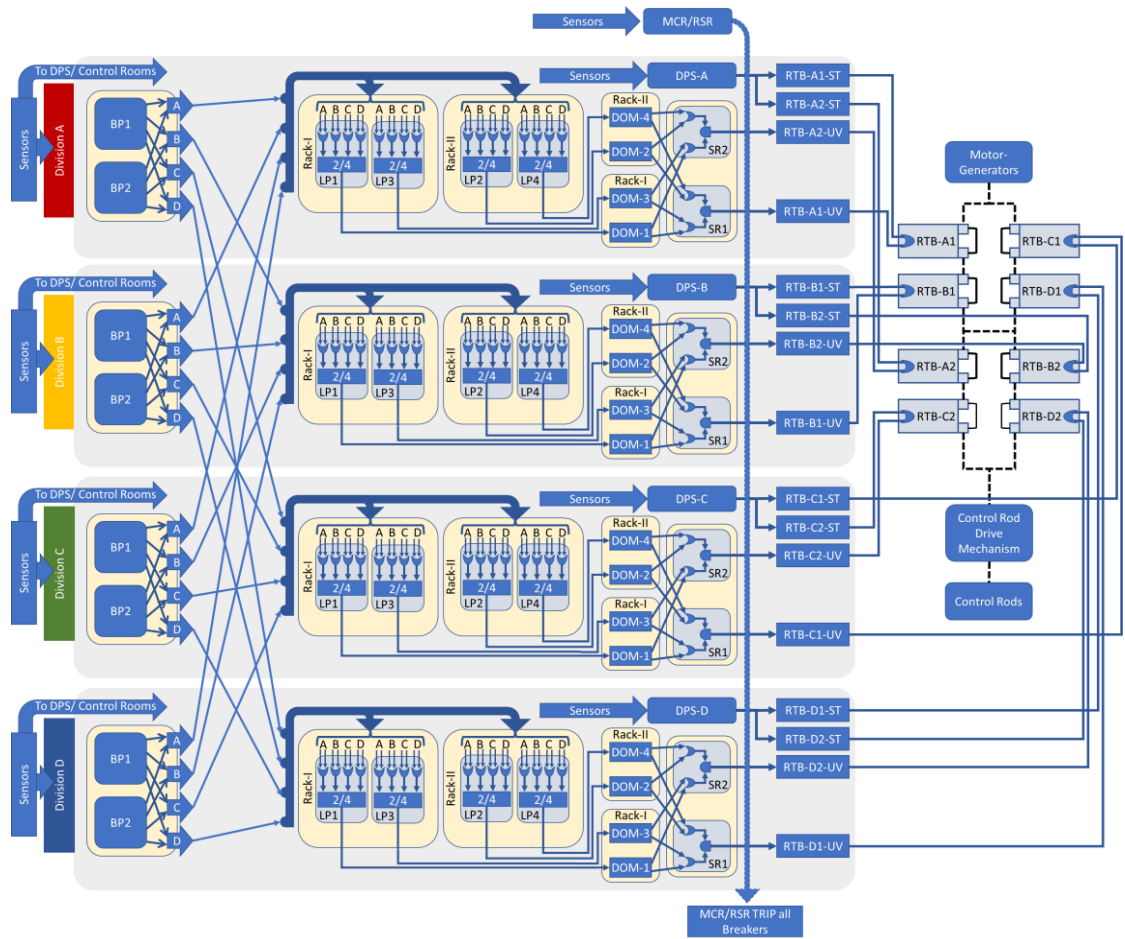

**Table 4. Total Hardware and Software Failure Probabilities for CCF Case Study.**

| Components | Hardware failure | Total Hardware failure probability | Software failure | Total Software failure probability |
|---|---|---|---|---|
| BPs | YES | 4.00E-5 | YES | 1.871E-4 |
| LPs | YES | 6.48E-5 | YES | 1.871E-4 |
| Digital Output Modules | YES | 1.64E-5 | N/A | N/A |
| Selective Relay | YES | 6.20E-6 | N/A | N/A |
| RTB-UV device | YES | 1.70E-3 | N/A | N/A |
| RTB-Shunt device | YES | 1.20E-4 | N/A | N/A |
| RTBs | YES | 4.50E-5 | N/A | N/A |
| All hardware values came from [27]. | | | | |

The details of the RTS were based on limited publicly available information [28], consequently some assumptions were made to complete the case study: (1) there is no diversity in the software; (2) all hardware components are not diverse (unless otherwise specified); (3) installation teams and maintenance teams are assumed identical for each CCCG; (4) each set of identical components that are part of the same CCCGs have the same total failure probabilities; (5) The software failure probability of the BPs were quantified[*] by the Bayesian and Human reliability analysis (HRA)-aided method for the reliability analysis of software (BAHAMAS). For convenience the failure probability of the BPs and LPs are assumed to be identical.

---

[*] The initial demonstration of BAHAMAS assumed a generic software component layout consisting of an input, an output, a central processing unit, and memory modules; each module was assumed to have software. The current work followed the same format given in the original publication, but assumes software is only found within the memory of each PLC processor.

The first step shown in Figure 2 is to assign the CCCGs after identifying the identical components and their coupling factors. There are eight identical BPs in the RTS, two per division. They each have an identical function and are assumed to share the same features, except for their installation location. All BPs share identical coupling factors, except for location, resulting in two CCCGs. One CCCG is based on shared function, hardware, software, and manufacturer. The second CCCG considers location. Table 5 shows the CCCGs identified for the BPs using the Redundancy-guided Systems-theoretic Hazard Analysis (RESHA) method developed at INL [29, 30]. Location creates an operational environment that is unique for software of the BPs. Despite having identical software, input from division-specific sensors creates the potential for the BPs to have division-specific CCFs associated with their operational conditions.

Table 5. CCCGs for the BPs.

| CCCGs | | Coupling Factors |
|---|---|---|
| 1 | All BPs | Function, Hardware, Software, & Manufacturer |
| 2 | Division A: BP1, BP2 | Location (Division A) |
| 3 | Division B: BP1, BP2 | Location (Division B) |
| 4 | Division C: BP1, BP2 | Location (Division C) |
| 5 | Division D: BP1, BP2 | Location (Division D) |

The next step from Figure 2 is to define the beta-factor parameters. Each CCCG receives a score for each sub-factor category. Sub-factors are scored according to the guidance provided in [21], with additional provisions for software as indicated in the preceding section. For example, CCCG1 for the BPs receives an A+ for Input Similarity. Specifically, CCCG1 consists of eight BPs (i.e., $m=8$). Each division receives its own sensor input that is shared by its BPs (i.e., $s=4$). The result is $R=s/m=0.5$ (i.e., A+ from Table 3). Table 6 shows the sub-factor scores for the BPs of CCCG1 and the calculation for beta based on Equation (15). The BPs for CCCGs 2–5 share the same qualitative features and receive beta factor scores of 0.123 and 0.568 for their hardware and software, respectively.

Table 6. Sub-Factor Scores for BPs CCCG 1 (All BPs CCF).

| Sub-factors | Hardware | | Software | |
|---|---|---|---|---|
| Redundancy (& Diversity) | B+ | 212 | A | 23976 |
| Separation/Input Similarity | E | 8 | A+ | 10112 |
| Understanding | A | 1800 | A | 7992 |
| Analysis | D | 25 | D | 45 |
| MMI | C | 173 | C | 379 |
| Safety Culture | E | 5 | E | 7 |
| Control | D | 25 | D | 28 |
| Tests | C | 69 | C | 379 |
| Beta for the CCCG | $\beta_{HD1} = 0.045$ | | $\beta_{SW1} = 0.429$ | |

The next step from the CCF modeling flowchart is to determine the CCFs. The BPs have multiple CCCGs; therefore, the modified BFM is used. For example, Division A, BP1 is found in two groups, CCCG1 and CCCG2, as shown in Table 5. Equations (7, 10 – 14) are used to find the independent and dependent failures of the BPs. The results of the CCF analysis are shown in Table 7. Note that RACK, DIVISION, and ALL correspond to the CCCG categories, while INDIVIDUAL corresponds to individual component failure. The CCCG ALL contains all the identical components within the system of interest. The given CCCG categories are not shared by all components; hence, there are no RACK CCCGs for the RTBs. Regarding the results, there is a difference between the software and hardware CCCGs of the LPs. The hardware CCCGs for the LPs are separated by location, just like the BPs. However, the potential for DIVISION and RACK level CCFs are precluded from consideration because there is nothing to distinguish them from the CCCGs representing all LPs; according to the case study, each LP has the same software and receives the same inputs. By contrast, the BPs have the potential for input variation amongst divisions. Thus, the BPs have DIVISION level software CCCGs, but the LPs do not. The results show that our methodology allows predicted software CCF to represent a larger failure probability than independent failure which matches our assumptions for a high redundancy low diversity software system.

Table 7. Hardware and Software Failure Probability for RTS Components.

| Component | INDIVIDUAL | RACK | DIVISION | ALL | Total |
|---|---|---|---|---|---|
| BPs-Hardware | 4.000E-05 | N/A | 5.943E-06 | 2.187E-06 | 4.813E-05 |
| LPs-Hardware | 6.480E-05 | 1.076E-05 | 7.647E-06 | 3.961E-06 | 8.717E-05 |
| DOMs | 1.640E-05 | 1.706E-06 | 1.015E-06 | 1.983E-07 | 1.932E-05 |
| Selective Relay | 6.200E-06 | N/A | 6.073E-07 | 7.059E-08 | 6.878E-06 |
| RTB-UV device | 1.700E-03 | N/A | N/A | 1.763E-05 | 1.718E-03 |
| RTB-Shunt device | 1.200E-04 | N/A | N/A | 1.244E-06 | 1.212E-04 |
| RTB RTSS2 | 4.500E-05 | N/A | N/A | 1.944E-06 | 4.694E-05 |
| BPs-Software | 5.591E-07 | N/A | 1.062E-04 | 8.030E-05 | 1.871E-04 |
| LPs-Software | 8.086E-05 | N/A | N/A | 1.062E-04 | 1.871E-04 |

## 4. CONCLUSION

This work introduces an approach for modeling software CCFs. A software CCF will be the result of a shared root cause (i.e., a trigger event and a latent fault) leading to the failure of two or more components by means of a coupling mechanism. Given a group of redundant software components, variations in their operating environments may lead to some, but not all, components failing together. Variations in the operational environment may result from differences in maintenance staff, input variables sources, and installation teams. These subtle differences may lead to unique combinations of CCFs. Thus, it is essential to consider software-based coupling mechanisms when assessing the potential for CCFs within a digital I&C system. When a group of components share coupling mechanisms, they form a CCCG. For most analyses, the components that belong to a CCCG do not belong to any other groups. This is because the components have no other coupling factors to share with components outside their existing group. When components can be grouped into multiple CCCGs (e.g., based on software operating environments), it becomes difficult to model their failure probabilities using conventional methods.

The chosen methodology employs the modified BFM and PBF-2 for modeling software CCFs by introducing modifications to PBF-2 for defining software-specific model parameters. The modified BFM was selected because it conveniently models components with multiple CCCGs. Normally, CCF methods rely on historical data or experience to define model parameters. However, limited data associated with novel designs requires a solution for quantifying model parameters. Innovations to PBF-2, together with the modified BFM, allow for a successful quantification process for the multiple CCCGs under a limited-data scenario. Several aspects of CCF modeling remain for future work. First, PBF-2 defines model parameters by considering the quality of a component's defenses against CCF. The method only considers eight sub-factors for assessing beta. There may yet be additional software-specific qualitative attributes to refine PBF-2. In addition, future research may provide an enumeration of software-specific coupling factors to aid the selection of software CCCGs. The modified BFM can also be improved. In its current form, the method, as with other ratio-based methods, is limited to similar components; future work may provide guidance for CCFs between non-identical components. In conclusion, the approach developed for this work provides a convenient means to quantify software CCF given a lack of operational and allow components to be part of multiple CCCGs simultaneously. Future collaborations with industry partners may afford our team the opportunity to investigate the data-sufficient scenario. In this case, there will be many opportunities to improve our models.

**Acknowledgements**

The research activities and achievements documented in this paper were funded by the U.S. DOE's Light Water Reactor Sustainability Program, Risk Informed Systems Analysis Pathway. This submitted manuscript was authored by a contractor of the U.S. Government under DOE Contract No. DE-AC07-